\documentclass[times]{aastex62}

\usepackage{placeins}  %for FLOATBARRIER on images & to control MOVING float figures 

\usepackage{rotating}    
\received{2016 February 6}
\revised{2016 May 17}
\accepted{2016 May 23}
\submitjournal{AJ}

\shorttitle{Temporal Evolution of Chromospheric Oscillations in Flaring Regions}
\shortauthors{Monsue et al.}
\begin{document}

\title{Temporal Evolution of Chromospheric Oscillations in Flaring Regions -- A Pilot Study}

\correspondingauthor{Teresa Monsue}
\email{teresa.monsue@vanderbilt.edu}

\author[0000-0003-3896-3059]{Teresa Monsue}
\affil{Department of Physics and Astronomy, Vanderbilt University, Nashville, TN 37235, USA}

\author{Frank Hill}
\affil{National Solar Observatory, 3665 Discovery Dr, Boulder, CO 80303}

\author{Keivan G. Stassun}
\affil{Department of Physics and Astronomy, Vanderbilt University, Nashville, TN 37235, USA}



\begin{abstract}

We have analyzed H$\alpha$ intensity images obtained at a 1 minute cadence with the Global Oscillation Network Group (GONG) system to investigate the properties of oscillations in the 0--8 mHz frequency band at the location and time of strong M- and X-class flares. For each of three sub-regions within two flaring active regions, we extracted time series from multiple distinct positions, including the flare core and quieter surrounding areas. The time series were analyzed with a moving power map analysis to examine power as a function of frequency and time. We find that, in the flare core of all three sub-regions, the low-frequency power ($\sim$1--2 mHz) is substantially enhanced immediately prior to and after the flare, and that power at all frequencies up to 8 mHz is depleted at flare maximum. This depletion is both frequency and time dependent, which probably reflects the changing depths visible during the flare in the bandpass of the filter. These variations are not observed outside the flare cores. The depletion may indicate that acoustic energy is being converted into thermal energy at flare maximum, while the low-frequency enhancement may arise from an instability in the chromosphere and provide an early warning of the flare onset. Dark lanes of reduced wave power are also visible in the power maps, which may arise from the interaction of the acoustic waves and the magnetic field.

\end{abstract}



%----------------------------------------------------------------------------------------
% 1.	INTRODUCTION
%----------------------------------------------------------------------------------------
\section{Introduction} \label{sec:intro}

\par
The relationship between solar acoustic oscillations, active regions, and energetic flares remains an open question in solar physics. A fundamental question is whether acoustic oscillations are enhanced, suppressed, or perhaps both, in active regions and by flares, and how these power enhancements and/or suppressions behave as functions of time before, during, and after energetic flaring events.  Solar flares release great amounts of energy and so in principle are capable of exciting acoustic oscillations in the magnetically active sunspot regions, perhaps by exciting velocity oscillations in regions where a higher-class solar flare has taken place \citep{Kumar2006}.  The first clear observations of helioseismic waves produced by a flare were by Kosovichev and Zharkova \citeyearpar{Kosovichev1998} using data from the Michelson Doppler Imager on board the Solar and Heliospheric Observatory space mission.  It is expected that this production of waves will change the characteristics, such as the power, of the p-modes \citep{Ambastha2002,Ambastha2003,Ambastha2003B,Kumar2006,Kumar2006I,Kumar2010,Kumar2011}.  

\par
In addition, Braun \emph{et al.} \citeyearpar{Braun1987} and others have observed 
that sunspots absorb acoustic power in the photosphere.  These authors found that outgoing waves were reduced in amplitude by 50\% compared to incoming waves \citep{Braun1987}.

\par
Observations of acoustic modes in the chromosphere using the H$\alpha$ line were first carried out by Elliott \citeyearpar{Elliott1969} and later by Harvey \citeyearpar{Harvey1993}.  Observations of H$\alpha$ intensity oscillations in solar flares were performed by \citet{Jain1999}.  Chromospheric oscillations in the 3.4 mHz and 5.6 mHz p-mode frequencies were found, confirming Elliott's study of their existence and also agreeing with \citet{Kneer1983}.  They surveyed 18 locations around two flares with Fourier power spectra and clearly found prominent 5 and 3 minute modes in H$\alpha$ \citep{Jain1998}.  Furthermore, the observation of short-time variations of the Sun's chromosphere observed in  H$\alpha$, contributes to our understanding of the atmospheric dynamics and could reveal progenitors for chromospheric heating mechanisms \citep{Kneer1985}.     

\par
We are interested in studying  the acoustic frequency spectrum to investigate  energy transfer in the chromosphere \citep{Kneer1983}.  Our method incorporates a short time-series analysis to study the  time variations in the Sun's chromosphere observed in H$\alpha$.  
Studying the chromosphere in the Fourier spectrum, and observing the temporal evolution gives us an advantage in that it provides a way to trace the temporal behavior of specific structures involved in the flare, which can be rather difficult to follow over the course of several minutes or hours otherwise
\citep{Kneer1985}. 

\par
A technique for studying the p-mode excitation due to solar flares in a three-dimensional Fourier analysis was devised by Jackiewiez and Balasubramaniam \citeyearpar{Jackiewicz2013} in their \emph{``frequency-filtered amplitude movies (FFAMs)"}.  This method is simple and powerful in that it preserves the initial 3-dimensional information in the inputted time series by incorporating a moving power map method.  Here we employ this novel approach to the Global Oscillation Network Group H$\alpha$ data set to create power-map movies (PMMs).  From these PMMs we investigate the H-alpha oscillatory modes across the frequency band (0 \textless \, $\nu$ \textless\, 8.33 mHz) and characterize the temporal and spatial evolution of these oscillations in flaring regions.  

\par
In Section~\ref{sec:data} we describe the X-ray flare data and the GONG H$\alpha$ time-series data that we employ in our analysis.  We specifically study three active regions, directly over sunspots, in which M- or X-class flares occurred on 2012 June 13 and 2012 July 12.  Section~\ref{sec:methods} describes our data reduction and analysis methods.  In Section~\ref{sec:results} we then present the results of how the frequency distribution evolves temporally and spatially by constructing a PMM of each region.  We find that, in the core regions of all three flares, the low-frequency power ($\sim$1--2 mHz) is substantially enhanced immediately prior to and after the flare, and that power at all frequencies up to 8 mHz is depleted at flare maximum. This depletion is both frequency- and time- dependent These variations are not observed outside the flaring region.  In addition, dark lanes of reduced wave power are also visible in the power maps.  Finally, in Section~\ref{sec:discussion} we conclude with a discussion of these results.  The observations may indicate that acoustic energy is being converted into thermal energy at flare maximum, while the low-frequency enhancement may arise from an instability in the chromosphere and provide an early warning of the flare onset.  We also suggest that the dark lanes observed may arise from the interaction of the acoustic waves and the magnetic field. Suggestions for next steps to advance this work are also provided.

%----------------------------------------------------------------------------------------
% 2.	OBSERVATIONAL DATA
%---------------------------------------------------------------------------------------- 
\section{Observational Data\label{sec:data}}

\par
The data set used here comprises GONG H$\alpha$ intensity images centered at a wavelength of 6562.8\textup{\AA}.  The GONG H$\alpha$ network became operational in 2010; it consists of a set of six detectors placed around the Earth for nearly continuous solar full-disk observations.  The images have a cadence of 1 minute and a format of 2048 $\times$ 2048 pixels.  

\par
Our initial study analyzes two individual data sets taken from two different GONG stations; El Tiede and Cerro Tololo.  Each data set covers approximately two hours of time (121 images) around two solar flare events that occurred on 2012 June 13 at 13:19UT and 2012 July 12 at 16:53UT (Table~\ref{table1}).  Each data set was centered in time on flare maximum, and started one hour before the event.  The June 13 flare had an NOAA classification of M1 in active region AR11504, and the July 12 flare was an X1.4 in active region AR11520.  The two detector data sets were not missing any time frame images.  

\par
We also use X-ray flux from the NOAA Space Weather Prediction Center (SWPC), obtained every minute with the Solar X-ray Imager (SXI) instrument on board the NOAA Geostationary Operational Environmental Satellite 15 (GOES-15) spacecraft. We compare the GONG time-series data with the X-ray flux in the 1-8\textup{\AA} wavelength band.

\par
\begin{figure}[htb!]
    \center
     \includegraphics[width=0.95\textwidth]{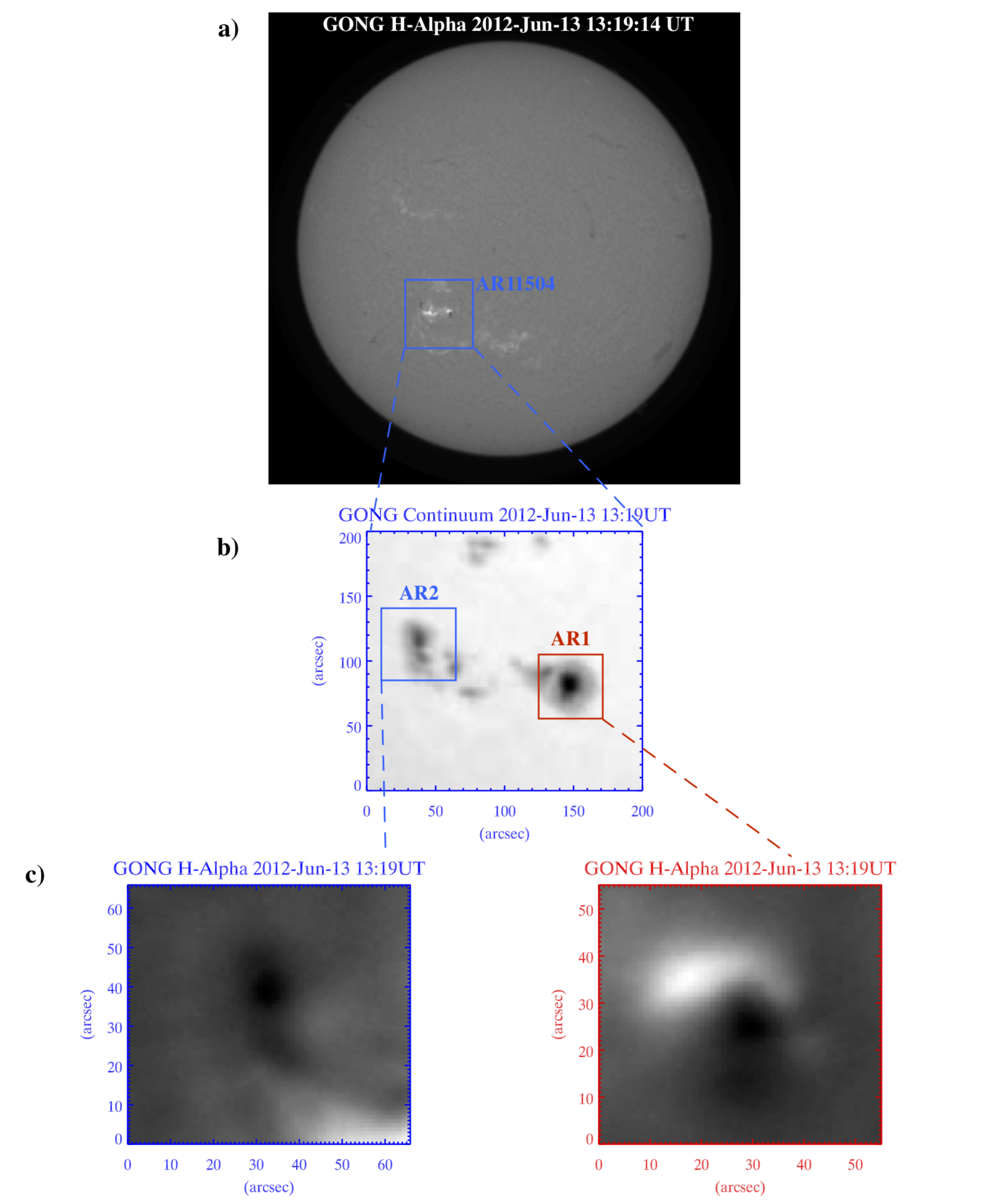}
 \caption{Analysis was done on solar flare regions directly over sunspots where the magnetic field was concentrated.  a)  The event is the M1 flare that occurred on 2012 June 13 at 13:19UT.  b)  Our region of interest is AR1, an ideal candidate with an active flaring region directly over a sunspot in AR11504, and the smaller active region AR2, a companion sunspot within that region.  c)  The main regions of interest have a rectangular area of approximately between 55$''$ $\times$ 55$''$ for AR1, and 65$''$ $\times$ 65$''$ for AR2.}
   \label{figure1}
      \end{figure}
\begin{figure}[!ht]
    \center
     \includegraphics[width=0.95\textwidth]{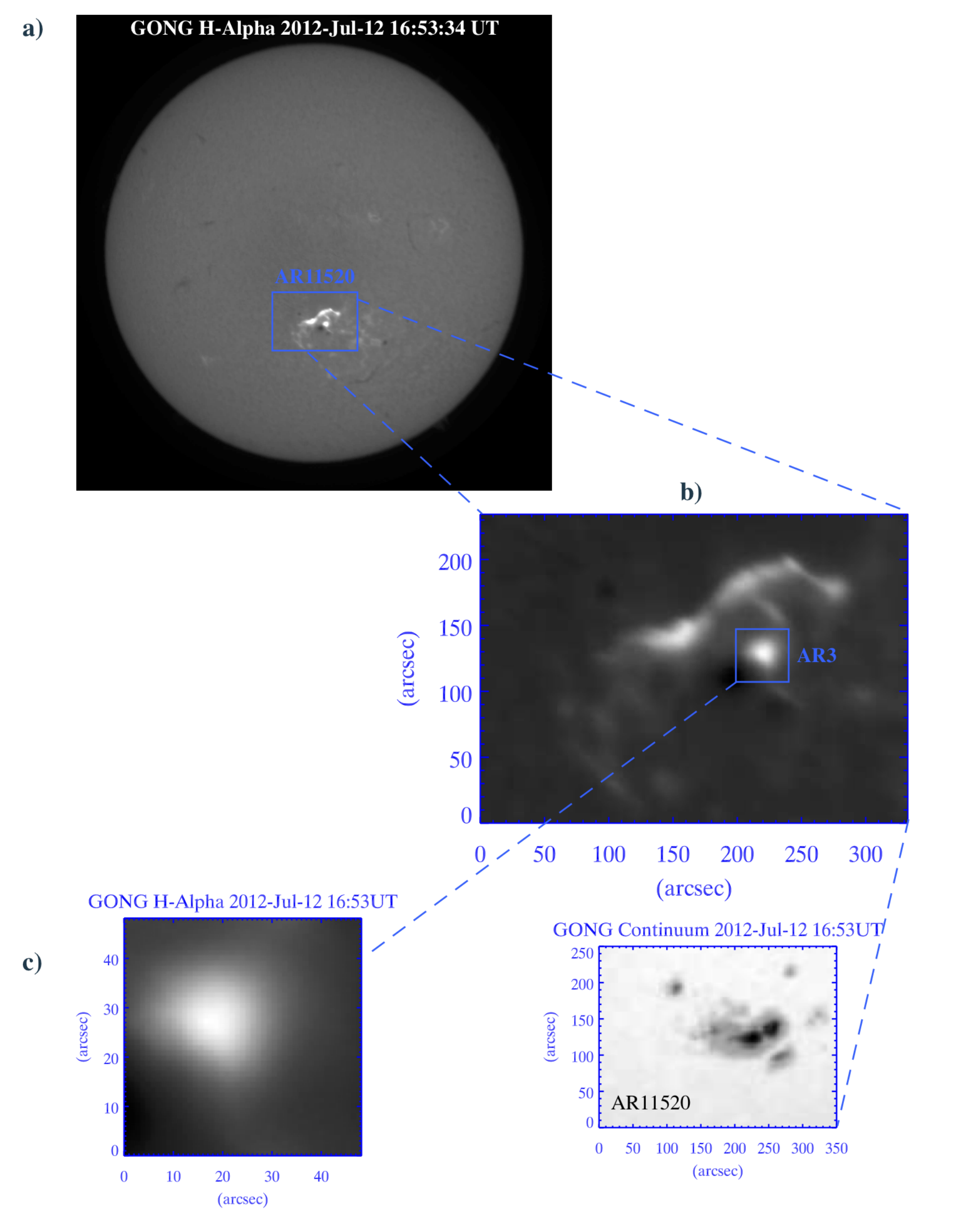}
     \caption{Analysis was done on solar flare regions directly over sunspots where the magnetic field was highly concentrated.  a) The event is the X1.4 flare that occurred on 2012 July 12 at 16:53UT.  b)  Our region of interest, AR3, was a nearly spherical flaring region directly over a sunspot in active region AR11520.  The sunspot morphology of AR11520 is depicted in the GONG continuum image below.  c) The main region of interest, AR3, has a rectangular area of approximately 48$''$ $\times$ 48$''$.}
\label{figure2}      
      \end{figure}

We carry out an analysis on active regions directly over sunspots during a flare.  There are a total of three regions of interest:  AR1 and AR2 in the June 13 flare (Figure~\ref{figure1}), and AR3 in the July 12 flare (Figure~\ref{figure2}).  The regions were chosen to be located close to the central meridian to reduce  projection effects \citep{Gizon2005}.  We restrict the GONG data to the same day of universal time (UT), and to one site for image stability. 

\newpage

%----------------------------------------------------------------------------------------
%	TABLE 1
%----------------------------------------------------------------------------------------
\begin{sidewaystable} % FOR SIDEWAYS TABLE

%\captionsetup{justification=centering,labelformat=simple,labelsep=colon}
\caption{Table of observational information of the three data sets.}
\centering
\label{T-observation}
\scalebox{.9}{
	\begin{tabular}{lccccccccc r@{.}l c} % define the column alignment
                                  % l: left, c: center, r: right
                                  % @{.} replace the inter-column by a .
\hline
\\ 
Data & Date & Active &Flare & AR & Event  & Images & GONG  &  Cadence \\
Set    &           & Region (AR) & Average Locations & Class & Time (UT) & & Station & [min] \\
\\     
\hline\hline %\hhline{=========}
 \\ 
AR1 & 13--June--2012& AR11504 & S17E26 (-399",-375") & M1& 13:19 & 121 & El Teide & 1\\
AR2 & 13--June--2012& AR11504 & S17E26 (-399",-375") & M1& 13:19 & 121 & El Teide & 1\\
AR3 & 12--July--2012& AR11520 & S17W08 (126",-385") & X1.4 & 16:53 & 121 & Cerro Tololo & 1\\

  \hline
	\end{tabular}
}
\label{table1}
\end{sidewaystable}
 \FloatBarrier      

%----------------------------------------------------------------------------------------
%	FIGURE 1
%----------------------------------------------------------------------------------------
 \FloatBarrier      
%%%%%%%%%%%%%%%%%%%%%%%%%%%%%%%%%%%%%%%%%%%%%%%%

%----------------------------------------------------------------------------------------
%	FIGURE 2
%----------------------------------------------------------------------------------------
 \FloatBarrier      
%%%%%%%%%%%%%%%%%%%%%%%%%%%%%%%%%%%%%%%%%%%%%%%%

%----------------------------------------------------------------------------------------
% 3.	DATA ANALYSIS AND REDUCTION METHODS
%----------------------------------------------------------------------------------------      
\section{DATA ANALYSIS AND REDUCTION METHODS\label{sec:methods}}

We employ a technique similar to the \emph{``FFAMs"} \citep{Jackiewicz2013} to construct a PMM. In summary, a PMM consists of a time series of acoustic power maps that show wave power as a function of frequency and space,  but with an initial starting time that is systematically offset. For a given starting time, the power maps are created by applying a fast Fourier transform (FFT) in the temporal direction for each pixel in the region of interest. The starting time is then shifted by one minute and the procedure is repeated.
 
\par
\indent
In detail, we start with a time series of GONG H$\alpha$ intensity images, $I(x,y,t)$ in a data cube with a cadence of 1 minute.  We define a region of interest ($I_{ROI}$)
%Equation~(\ref{equation1}), 
covering the area $x_1 \le x \le x_2$  
and $y_1 \le y \le y_2$ within the large data cube.  
 We extract the time series at each ROI pixel over a temporal length $N$ starting at time \emph{t\textsubscript{i}}.  Missing images are replaced by a value of zero.  The input data for the PMM are constructed by incrementing $t_i$. Explicitly, we have
%----------------------------------------------------------------------------------------
%	EQUATION 1
%----------------------------------------------------------------------------------------
\begin{equation}      %EQUATION 1
\label{equation1}
I_{\emph{ROI}}(x',y',t',t_i) = I(x_1 \le x \le x_2, \ y_1 \le y \le y_2,  \ t_i \le t \le t_i + N)
\end{equation}
For our analysis here, $t_i$ is incremented  by one minute, $0 \le t_i \le 60$, and $N = 60$, producing a set of 61 time series, each an hour long, that start at one hour prior to flare maximum, and end at one hour after flare maximum.
\par
\indent
To create a single frame, $P$, in the PMM, we apply an FFT in the temporal direction to the time series for each spatial pixel in $I_{ROI}$, and then take its modulus, producing
%----------------------------------------------------------------------------------------
%	EQUATION 2
%----------------------------------------------------------------------------------------
\begin{equation}               %EQUATION 2 
\label{equation2}
 P(x',y',\nu, t_i) = | FFT (I_{\emph{ROI}}(x',y',t',t_i)) | ^2
\end{equation}
where $\nu$ is the temporal frequency. The power $P$ is then averaged in frequency bins $\nu_i$, producing one frame in the PMM, Equation~(\ref{equation3}).
%----------------------------------------------------------------------------------------
%	EQUATION 3
%----------------------------------------------------------------------------------------
\begin{equation}                %EQUATION 3
\label{equation3}
PMM(x', y', \nu_j, t_i) = {{\Delta \nu} \over  {\nu_2 - \nu_1}}
\sum_{\nu_1 < \nu_j < \nu_2} P(x', y', \nu, t_i)
\end{equation}
where $\Delta \nu$ is the frequency resolution of the power spectrum (277.8 $\mu$Hz) and $\nu_1$ and $\nu_2$ are the lower and upper limits respectively of frequency bin $j$. We define a set of seven bins $\nu_j$ of width 1 mHz starting at 1 mHz, with the highest band covering 7--8.33 mHz.  

\par
\indent
Within the areas AR1--AR3, we extract a number of smaller regions of 3 $\times$ 3  pixels, or approximately 3.2$''$ in length, and then average the PMM values within these small areas.  Averaging the pixels over a smaller region improves the signal-to-noise ratio and allows us to isolate different physical conditions.  We kept all subregions the same size to maintain consistency within the experiment.

\par
\begin{figure}[h]%[htb]
    \center
     \includegraphics[width=\textwidth]{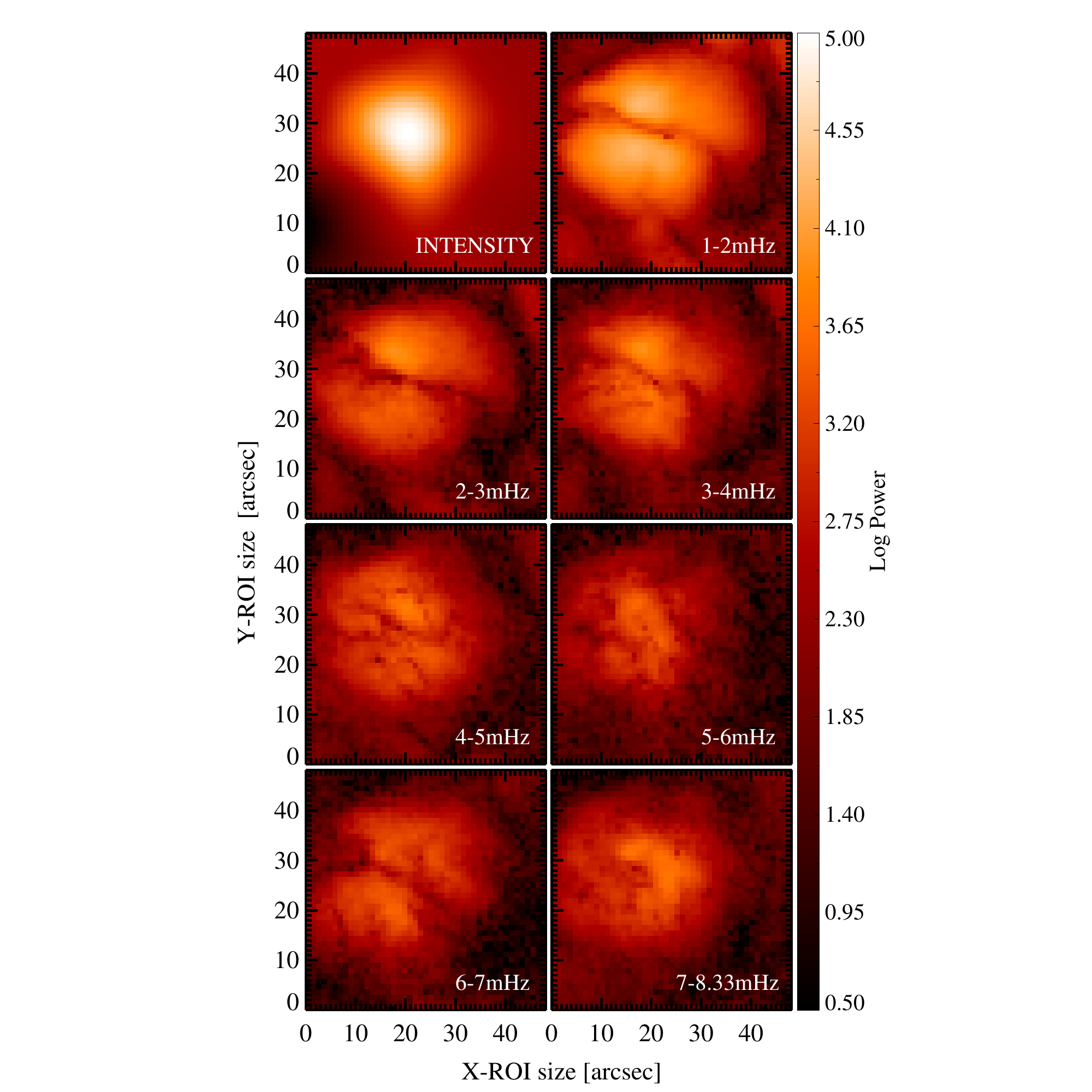}
     \caption{The flare intensity  and   frequency-binned PMM frames of the X1.4 flare on 2012 July 12 in region AR3.   The PMM frames are in the time window of 32--92 minutes.   Dark lanes are seen in each frame and are  sharper in the 1--2 mHz and 2--3 mHz ranges.  The power is measured in arbitrary instrumental units.}
 \label{figure3}
 \end{figure}

To compare the PMM data with the intensity, we carry out a 60 minute running average of the intensity that is incremented by one minute to provide the same sampling as the PMM. Figure~\ref{figure3} shows a power-map frame for region AR3 for each $\nu_j$ along with the averaged intensity.  The sampling window in Figure~\ref{figure3} is from 32 to 92 minutes in the total time series. 

%----------------------------------------------------------------------------------------
%	FIGURE 3
%----------------------------------------------------------------------------------------
 \FloatBarrier      
%%%%%%%%%%%%%%%%%%%%%%%%%%%%%%%%%%%%%%%%%%%%%%%%

%----------------------------------------------------------------------------------------
% 4.	RESULTS
%----------------------------------------------------------------------------------------
\section{RESULTS\label{sec:results}}

\par
In this section, we present the results of our time-series power spectrum analysis on each of the three flaring active regions defined in Section~\ref{sec:data}.  Within each of the three main analysis regions---AR1, AR2, and AR3---we define several subregions in and around the centers of the active regions in order to relate the temporal behavior of the oscillation power to different spatial positions.  As we discuss below, the overall results for all three analysis regions are qualitatively similar, so we describe the results for one of the three regions in detail and then more briefly summarize the similar results for the other two regions.
We find that, in the core regions of all three flares, the low-frequency power ($\sim$1--2 mHz) is substantially enhanced immediately prior to and after the flare, and that power at all frequencies up to 8 mHz is depleted at flare maximum. This depletion is both frequency- and time-dependent. We also observe dark lane features in the power maps. We defer a discussion and interpretation of these general findings to Section~\ref{sec:discussion}.

%----------------------------------------------------------------------------------------
%	AR3 SECTION
%----------------------------------------------------------------------------------------
\subsection{AR3} %%%%%%%%%%%%%%
  \label{S-AR3}

\par
\begin{figure}[h]%[htb]
       \centering
    \includegraphics[width=\textwidth]{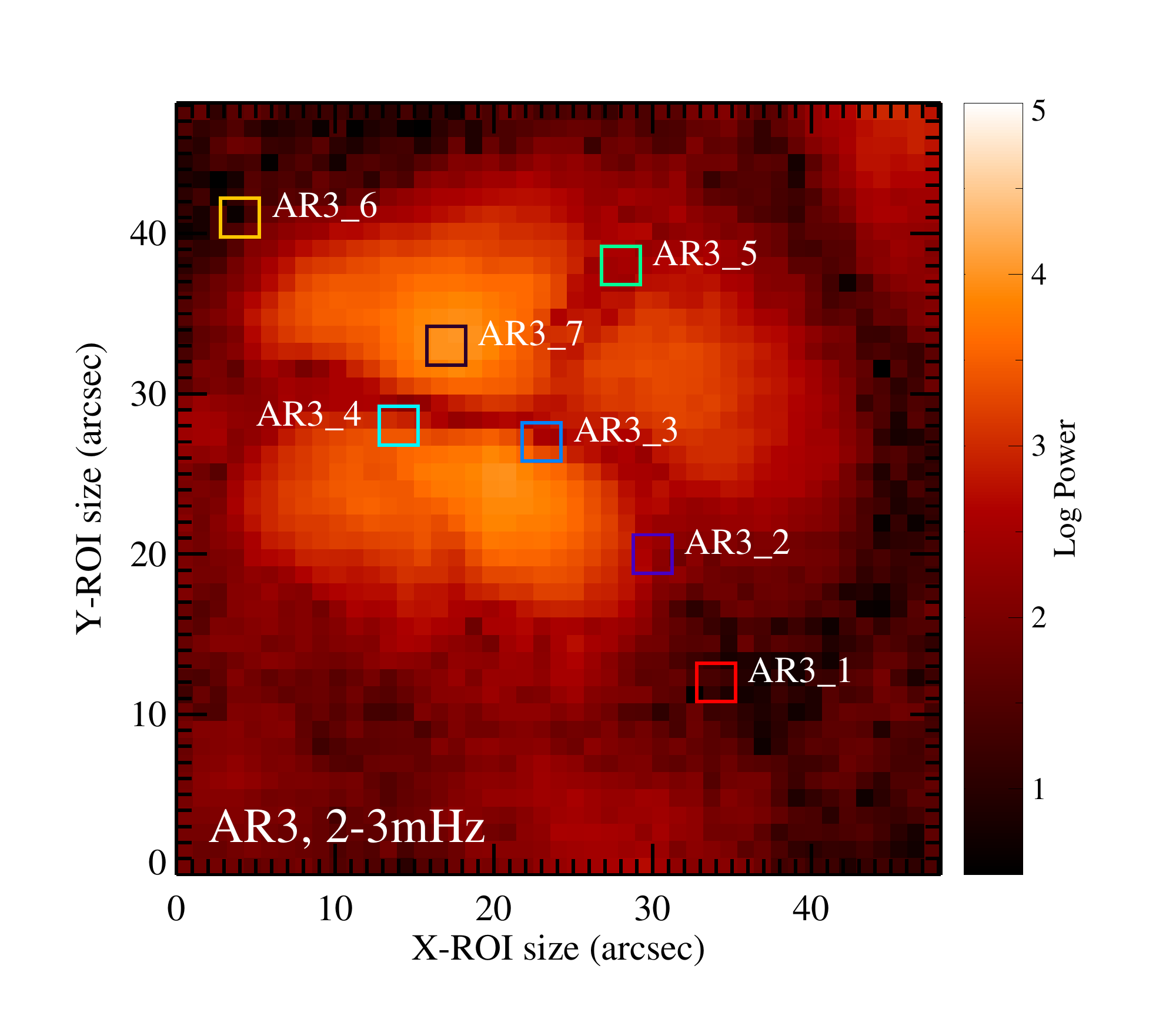}

\caption{Region AR3  has a rectangular area of  approximately 48.15$''$ $\times$ 48.15$''$. There are seven subregions sampled in AR3 for the X1 solar flare on 2012 July 12.  The subregions are approximately 3.2$''$ $\times$ 3.2$''$ in size.  In the above figure representing AR3, the PMM frame depicted is in the time window of 34--94 minutes.}      
\label{figure4}
\end{figure}

There are three main groups of subregions for area AR3.  Figure~\ref{figure4} shows seven rectangular subregions, each approximately 3.2$''$ $\times$ 3.2$''$, distributed around the solar flare and  the dark lanes.  The subregions are selected to probe the temporal behavior at various levels of overall power and at various positions relative to the center of the active region. We present the results by grouping the subregions accordingly. Obviously these are not the only subregions that could be selected; we emphasize that this selection of subregions is made arbitrarily, and on the basis of visual impression, but in an attempt to sample the active region at various representative locations relative to the peak of activity.  The regions with the lowest  power (log power 0.5 to 2.0), are AR3\_1 and AR3\_6 (Figure~\ref{figure7} (a) and (b)).  The  regions with mid-level  power levels (log power 2.0 to 3.0) are AR3\_2 and AR3\_5 (Figure~\ref{figure6} (a) and (b)).  The regions located in the middle of the flare (AR3\_3, AR3\_4 and AR3\_7) exhibited the greatest amount of  power (Figure~\ref{figure5}(a)--(c)), with logarithmic values from 3.0 to 4.5.

%----------------------------------------------------------------------------------------
%	FIGURE 4
%----------------------------------------------------------------------------------------
 \FloatBarrier      
 %%%%%%%%%%%%%%%%%%%%%%%%%%%%%%%%%%%%%%%%%%%%%%%%

\subsubsection{Inner Flaring Regions -- \emph{Locations 3, 4, and 7 in AR3}}

\par
\begin{figure}[h]%[htb]
\centering
\includegraphics{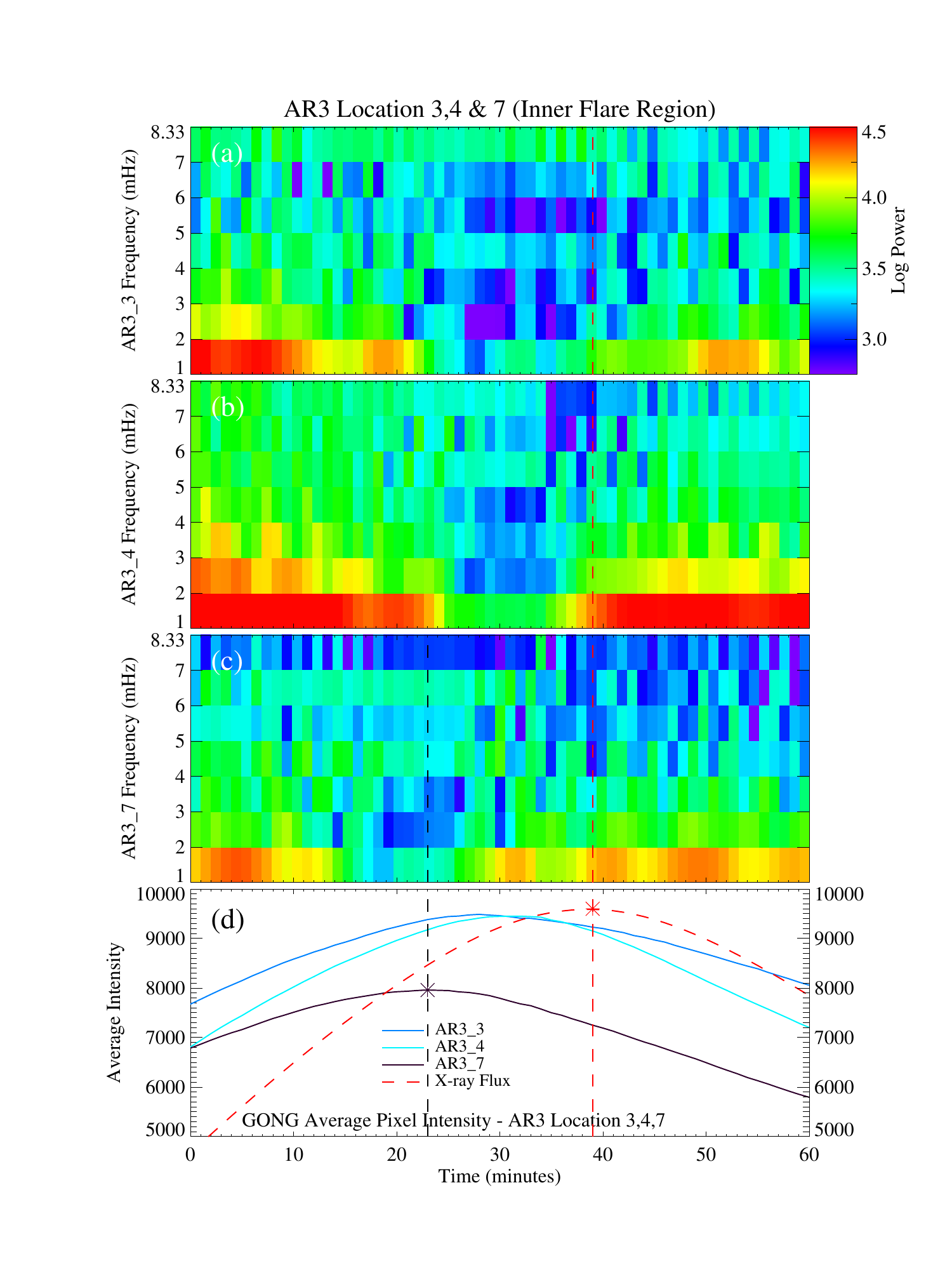} % NEW PLOT
 \caption{Time-frequency power plots of the three subregions AR3\_3 (a), AR3\_4 (b) and AR3\_7 (c) along with the corresponding average intensity (d).  The \emph{GOES} X-ray flux is scaled as a reference for the overall solar flare event.  The red dashed line indicates the time of flare maximum in the time series at 39 minutes.}
 \label{figure5}
      \end{figure}

These three subregions are placed along the dark lanes and in a bright region in the flare.   Figure~\ref{figure5}(a)--(c) shows the power as a function of time and frequency in these subregions. We observe an increase in power across the entire frequency band once the flare begins.  However, as the intensity increases in each subregion, we observe a suppression of power that begins first at the higher frequencies, and then moves toward lower frequencies as time progresses. Maximum power suppression at all frequencies occurs at the time of the maximum local intensity. The power then increases back toward pre-flare levels, with the lower frequencies recovering first. The result is the appearance of a ``V''-shaped feature in the color plots of Figure~\ref{figure5}. For each subregion,
the lowest frequency band (1--2 mHz) shows the greatest power. 

\par
The overall temporal extent of the ``V''-shaped decrease in power is shortest for subregion AR3\_4, which might be a consequence of the narrow spatial width of the dark lane in that region. The suppression in power in AR3\_3 (Figure~\ref{figure5}(a)) is wider in time, perhaps due to the larger width of the dark lane in that area.  AR3\_7 (Figure~\ref{figure5}(c)), free of dark lanes, shows the longest time period of power suppression.
Figure~\ref{figure5}(d) shows the total average intensity variations for each of the three regions along with the \emph{GOES} X-ray flux.

%----------------------------------------------------------------------------------------
%	FIGURE 5
%----------------------------------------------------------------------------------------
 \FloatBarrier                     
%%%%%%%%%%%%%%%%%%%%%%%%%%%%%%%%%%%%%%%%%%%%%%%%

\newpage
\subsubsection{Outer Flaring Regions -- \emph{Locations 2 and 5 in AR3}} 

\par
\begin{figure}[h]%[htb]
       \centering
           \includegraphics{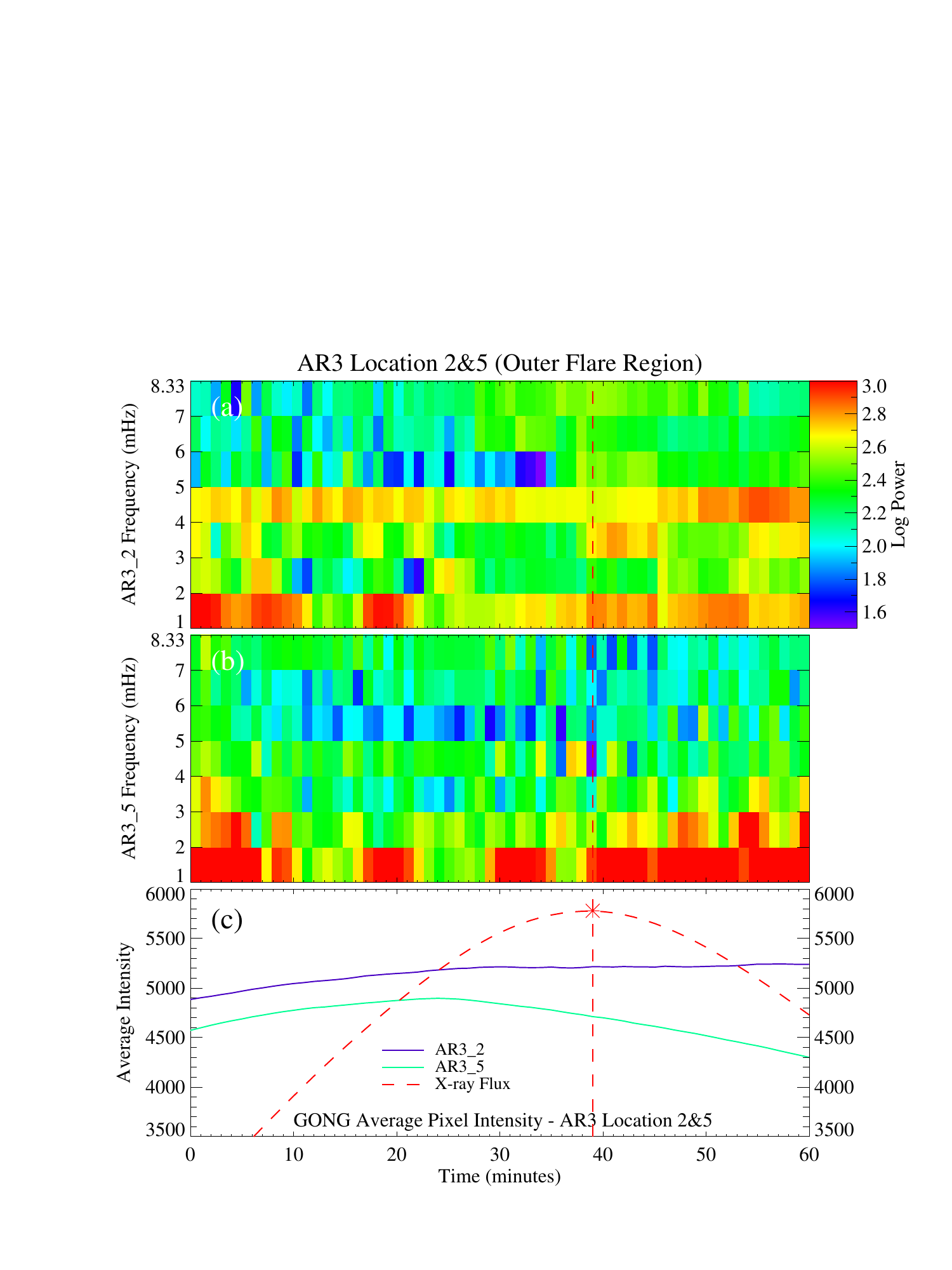}

 \caption{Analysis of the two outer flare regions, AR3\_2 (a) and AR3\_5 (b), along with the corresponding pixel intensity and \emph{GOES} X-ray flux (c).  The \emph{GOES} X-ray flux is scaled as a reference for the overall solar flare event.  The red dashed line indicates the solar flare event in the time series at 39 minutes.}
 \label{figure6}
      \end{figure}

Regions AR3\_2 and AR3\_5 both lie on the outer edge of the dark lanes, as shown in Figure~\ref{figure4}.  The overall power for these two regions are in the middle range of log power, 2.0--3.0. The corresponding time-frequency plots are shown in Figure~\ref{figure6}.  

\par
These plots have a substantially different qualitative nature compared to those in Figure~\ref{figure5}. Here, there is no sign of the ``V'' shape in these time-frequency images. For region AR3\_2 (Figure~\ref{figure6}(a))  there is an overall increase in power for the 4--5 mHz frequency band, with logarithmic values in the range of 2.6 to 3.0 - also observed in region AR3\_1 (Figure~\ref{figure7}(a)).  A decrease in power for   frequencies above 5 mHz (Figure~\ref{figure6}(a)) is present at the start of the time series and persists to around 35 minutes.  There appears to be a trend of increasing power late in the time series  in the frequency band of 1--2 mHz.  For region AR3\_5 (Figure~\ref{figure6}(b)) there are three areas of power suppression within the 1--2 mHz band.  A period of fluctuating suppression is also observed in the 5--6 mHz band.

\par
Figure~\ref{figure6} shows that these two subregions have substantially lower levels of average intensity variations than those in Figure~\ref{figure5}. Thus suggests that rapid changes in the average intensity are at least partially responsible for the power variations in the oscillations.

%----------------------------------------------------------------------------------------
%	FIGURE 6
%----------------------------------------------------------------------------------------
 \FloatBarrier                    
%%%%%%%%%%%%%%%%%%%%%%%%%%%%%%%%%%%%%%%%%%%%%%%%

\subsubsection{Quiescent Regions -- \emph{Locations 1 and 6 in AR3}}

\par
\begin{figure}[h]%[htb]
       \centering
           \includegraphics{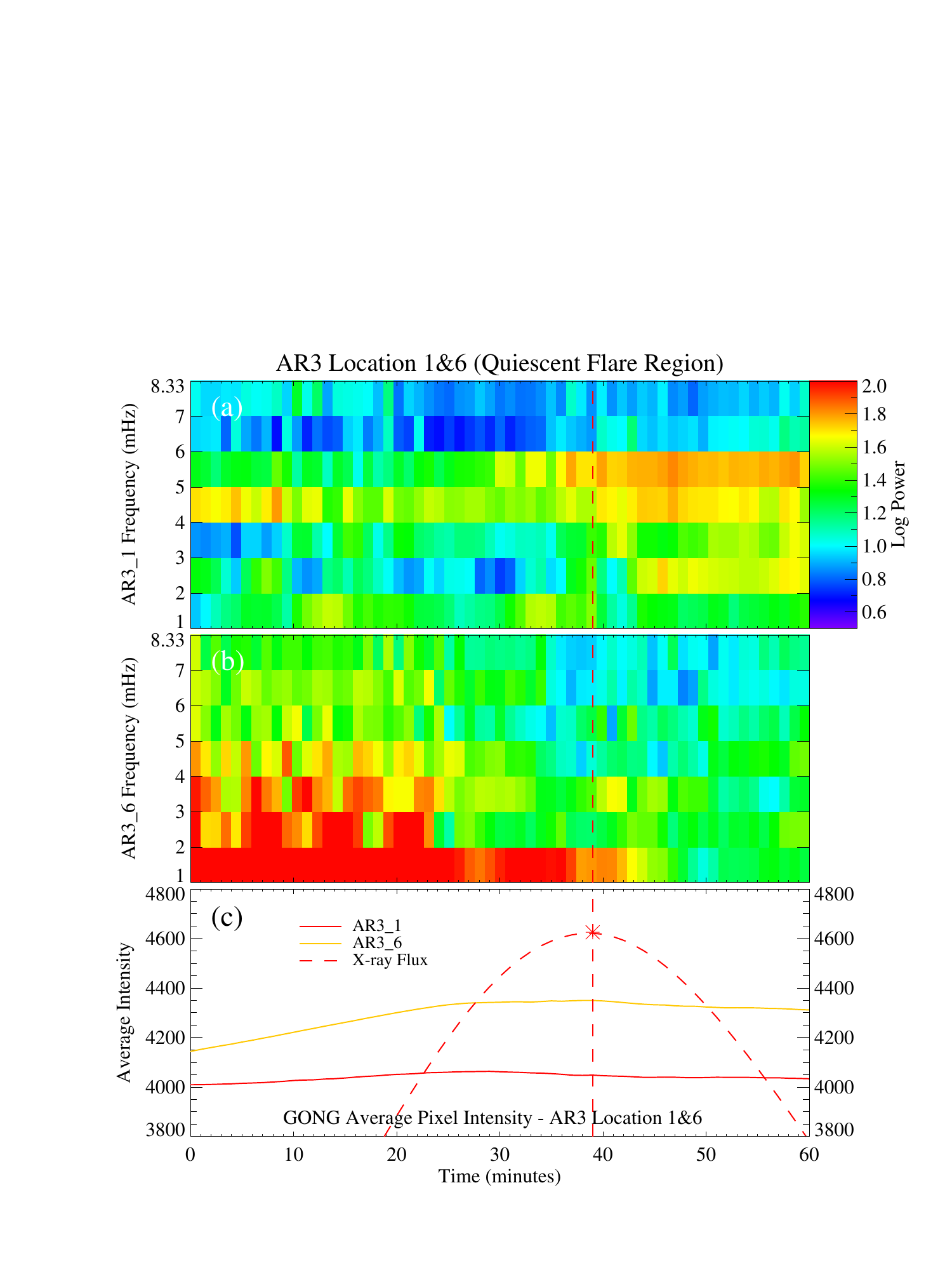}

 \caption{Analysis of the two quiescent flare regions: AR3\_1 (a) and AR3\_6 (b), along with the corresponding pixel intensity and \emph{GOES} X-ray flux (c).  The \emph{GOES} X-ray flux is scaled as a reference for the overall solar flare event.  The red dashed line indicates the solar flare event in the time series at 39 minutes.}
 \label{figure7}
      \end{figure}

Regions AR3\_1 and AR3\_6 both lie on the outer edge of the solar flare and in very dark outer regions.  The overall logarithmic power for these two regions is in the lowest range of 0.6--2.0, Figure~\ref{figure7}.  

\par
These areas also do not show the ``V''-shape suppression feature.  For region AR3\_1 (Figure~\ref{figure7}(a))  there is an overall increase in the power in the frequency band of 1--6 mHz.  Some power suppression is apparent in the time period of about 8--42 minutes. The overall power has a maximum in the 5--6 mHz band.  

\par
Region AR3\_6 (Figure~\ref{figure7}(b)) shows an overall  suppression of power in all frequency bands that starts at the time of maximum X-ray intensity. This region also shows a substantially higher power level at frequencies of 1--3 mHz than at 6--8 mHz.

\par
As in Figure~\ref{figure6}(c), the curves of average intensity curves in Figure~\ref{figure7}(c) do not show much relative variation. This is consistent with the hypothesis that the ``V'' shape is related to the presence of strongly varying intensity.

%----------------------------------------------------------------------------------------
%	FIGURE 7
%----------------------------------------------------------------------------------------
 \FloatBarrier                    
%%%%%%%%%%%%%%%%%%%%%%%%%%%%%%%%%%%%%%%%%%%%%%%%

%----------------------------------------------------------------------------------------
%	AR1 SECTION
%----------------------------------------------------------------------------------------
\subsection{AR1} %%%%%%%%%%%%%%
  \label{S-AR1}

\par
\begin{figure}[h]%[htb]
       \centering
           \includegraphics{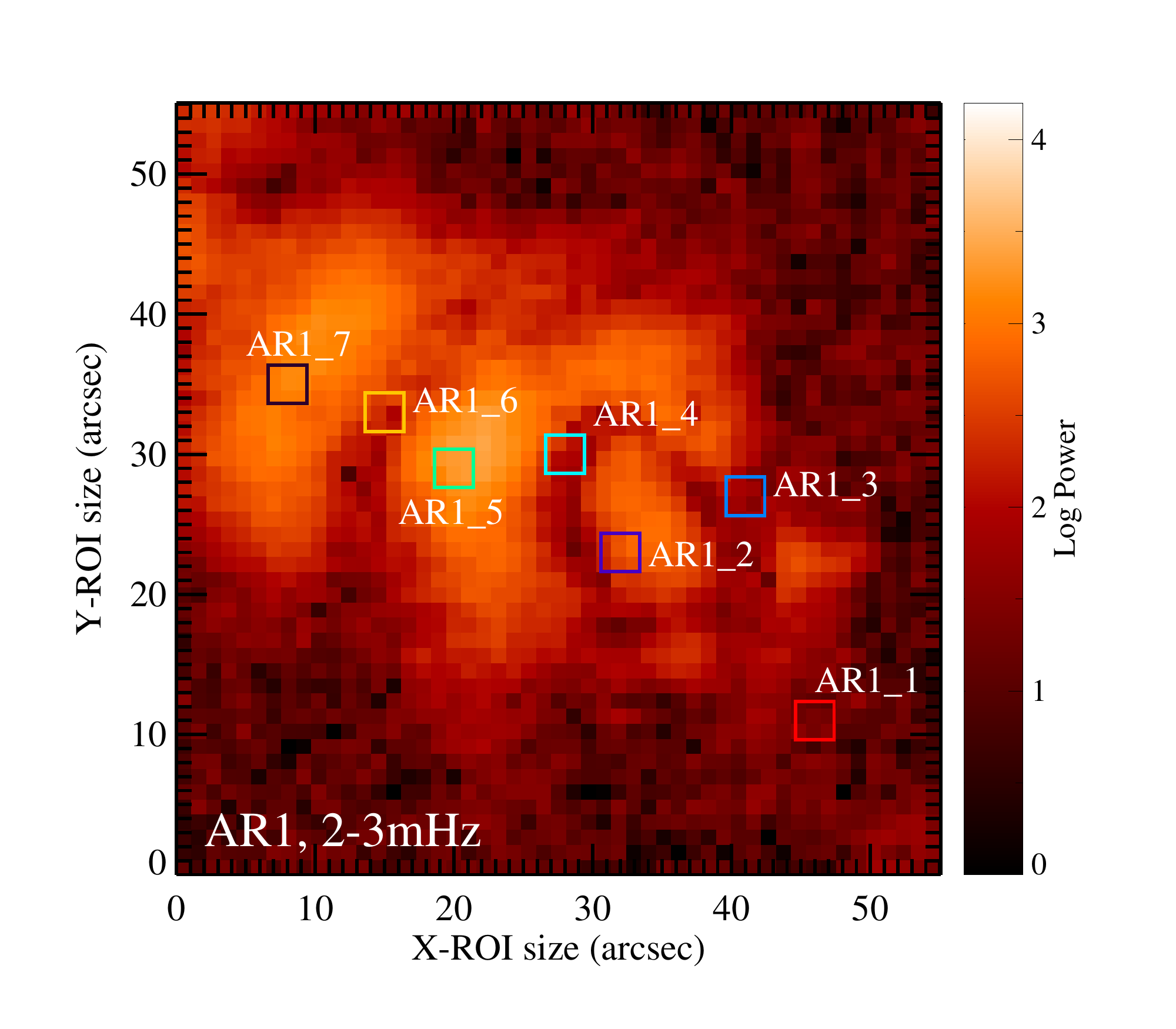}

\caption{AR1  has an area of  approximately 55$''$ $\times$ 55$''$. There are seven subregions  in AR1 for the M1 solar flare occurring on 2012 June 13.  In this figure, the PMM frame depicted is in the time window of 32 to 92 minutes.}     
\label{figure8}
\end{figure}
\begin{figure}[ht]%[htb]
       \centering
       \includegraphics{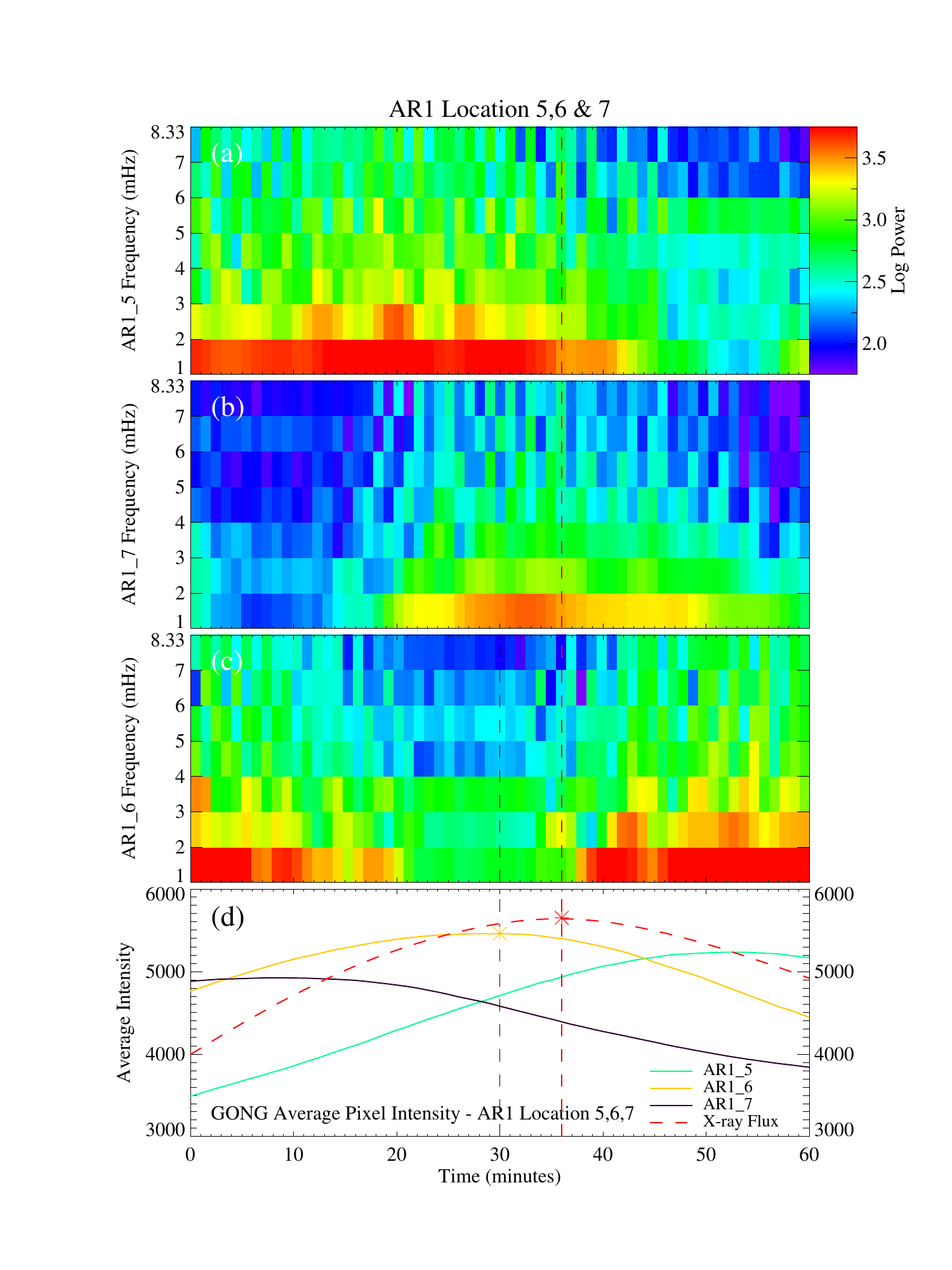} % NEW PLOT

 \caption{Time-frequency  power plots for regions AR1\_5, AR1\_6, and AR1\_7, showing the suppression of power in plots (a)--(c), correlating with the peak intensity (d).  All three regions exhibit a peak in power around 3.75 at the lowest frequency bands of 1--2 mHz and then gradually decreasing as the frequency bands increase.  The \emph{GOES} X-ray flux is scaled as a reference for the overall solar flare event in the time series at 36 minutes.}

     \label{figure9}
      \end{figure}

Region AR1 has an area of approximately 55$''$ $\times$ 55$''$ (Figure~\ref{figure8}).  As with AR3, we select seven subregions intended to probe the temporal evolution of the flaring event at various spatial locations relative to the center of the active region.  Each of the seven subregions samples a rectangular area with 3.2$''$ per side.  AR1 displays similar behavior to that of AR3 and so only the most significant of results (Figure~\ref{figure9}) will be discussed.  We present the results for locations AR1\_5, AR1\_6 and AR1\_7, the regions where the M1 flare on 2012 June 13 was the most intense.

%----------------------------------------------------------------------------------------
%	FIGURE 8
%----------------------------------------------------------------------------------------
\FloatBarrier      
 %%%%%%%%%%%%%%%%%%%%%%%%%%%%%%%%%%%%%%%%%%%%%%%%

\newpage
\par
The subregions where the M1 flare in AR1 was the strongest were AR1\_5, AR1\_6, and AR1\_7.  The relationship between the peak of the intensity and the suppression of power is clearly observed in Figure~\ref{figure9}.  The X-ray flux of the flare is scaled and plotted along with the intensity to show the time of maximum  overall flaring activity (Figures~\ref{figure9}(d)).  
The suppression of power is observed as a ``V'' shape for locations AR1\_5 and AR1\_6 (Figure~\ref{figure9}(a) and (b)) with results similar to AR3\_3, AR3\_4 and AR3\_7 (Figure~\ref{figure5}). Again we see a relatively large variation in average intensity, with the point of the ``V'' corresponding to the intensity maximum in Figure~\ref{figure9}(d).

\par
Figure~\ref{figure9}(c), shows a rather different situation with a clear indication that the power is suppressed not only at the peak of the intensity but also near a minimum. This will be discussed later.

\par
Once again, as seen in almost all of our time-frequency power plots, the power is greatest at low frequencies below 2 mHz.

%----------------------------------------------------------------------------------------
%	FIGURE 9
%----------------------------------------------------------------------------------------
 \FloatBarrier     
 %%%%%%%%%%%%%%%%%%%%%%%%%%%%%%%%%%%%%%%%%%%%%%%%

%----------------------------------------------------------------------------------------
%	AR2 SECTION
%----------------------------------------------------------------------------------------
\subsection{AR2} %%%%%%%%%%%%%%
  \label{S-AR2}

\par
\begin{figure}[h]%[htb]
       \centering
           \includegraphics{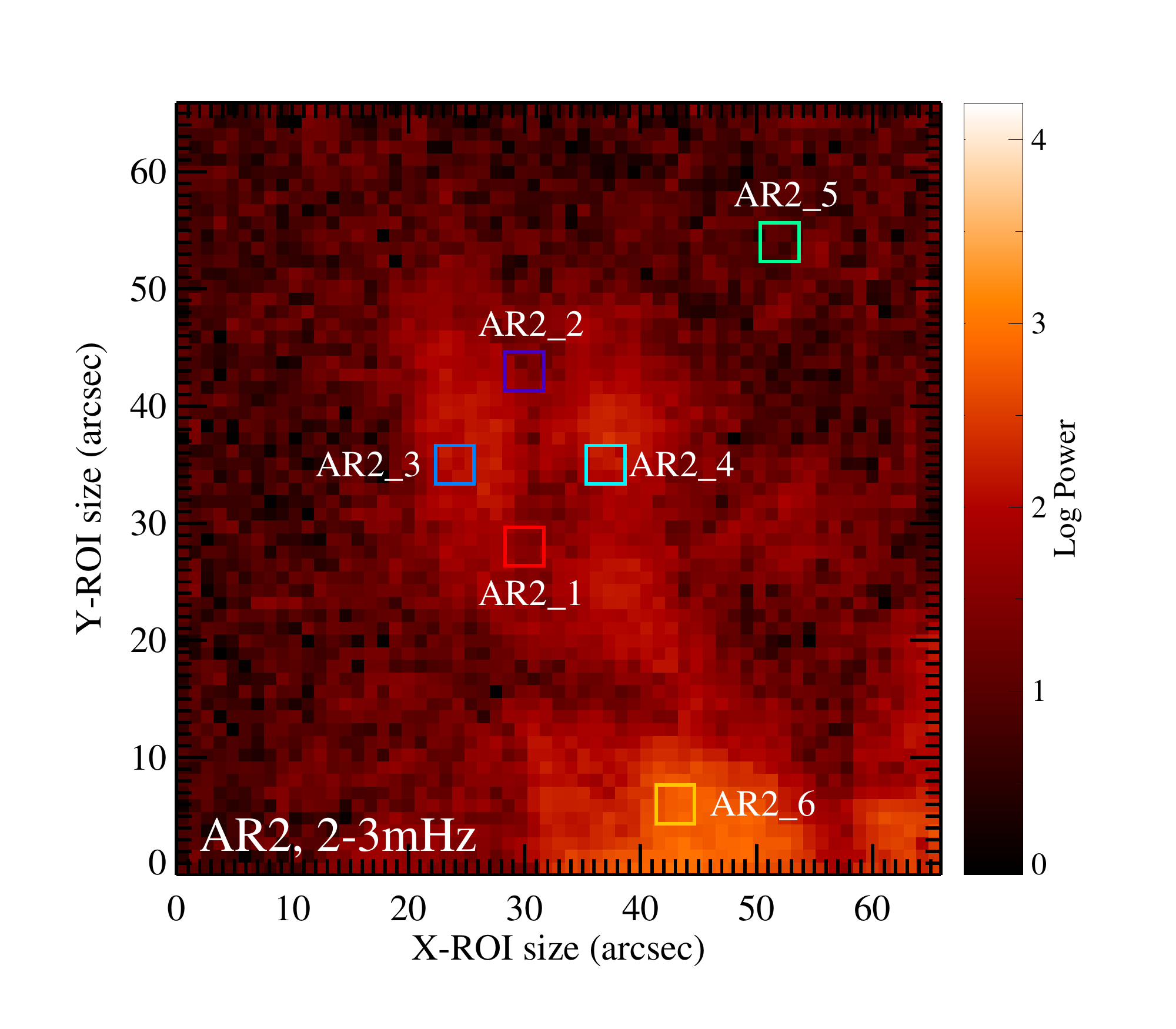}

\caption{AR2 has a rectangular area of approximately 66$''$ $\times$ 66$''$. There are six subregions in total in AR2 for the M1 solar flare on 2012 June 13.  In this figure, the PMM frame depicted is in the time window of 32--92 minutes.}     
\label{figure10}

\end{figure}
\begin{figure}[ht]%[htb]
       \centering
           \includegraphics{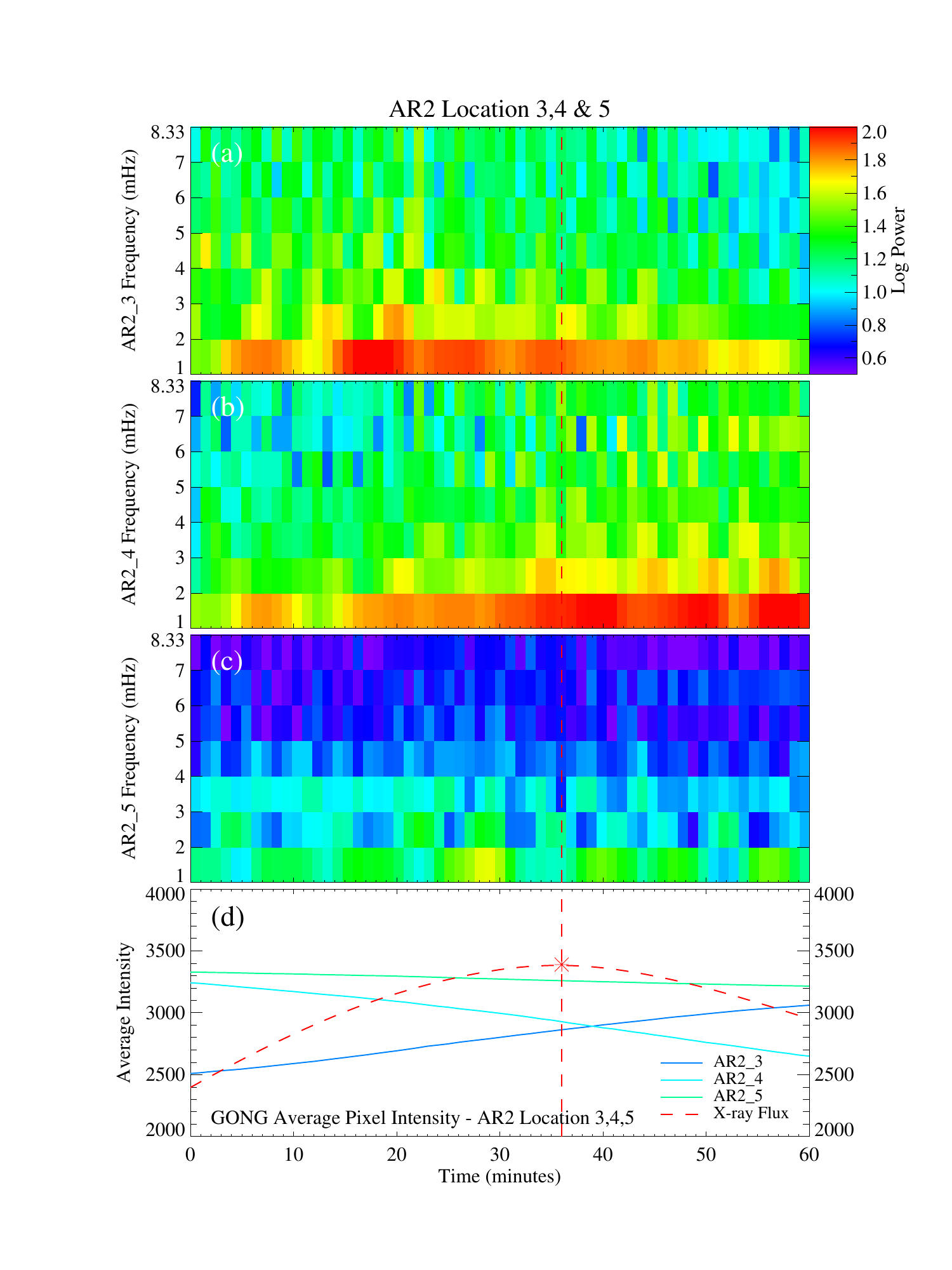}

 \caption{Acoustic power observations for regions AR2\_3, AR2\_4, and AR2\_5, showing the suppression of power in plots (a) and (b) and a quiet region for reference (c).  The \emph{GOES} X-ray flux is scaled as a reference for the overall solar flare event in the time series at 36 minutes.}
  
     \label{figure11}
      \end{figure}

Region AR2 is approximately 66$''$ $\times$ 66$''$ in size (Figure~\ref{figure10}).  This was the smaller sunspot region involved in the M1 solar flare on 2012 June 13 (Figure~\ref{figure1}).  Once again, we defined several subregions in order to probe the temporal behavior of the activity at various spatial positions relative to the visual center of the activity.  Six  subregions  of 3.2$''$ in size were analyzed.  AR2 displayed results similar to those of both AR1 and AR3.  Here we present an analysis of two of the subregions close to the center of the active region (Figure~\ref{figure11}, (a) and (b)), and a control quiet subregion for comparison (Figure~\ref{figure11}(c)).

%----------------------------------------------------------------------------------------
\vspace{-0.25cm} %Reduce space to bring up the section below
%----------------------------------------------------------------------------------------
%----------------------------------------------------------------------------------------
%	FIGURE 10
%----------------------------------------------------------------------------------------
 \FloatBarrier      
 %%%%%%%%%%%%%%%%%%%%%%%%%%%%%%%%%%%%%%%%%%%%%%%%

\par
Region AR2 provides observations of the oscillatory power behavior when the average intensity is slowly varying.  In Figure~\ref{figure11}(d) the intensity  for AR2\_3 is slowly increasing, while it is decreasing for  AR2\_4. The corresponding time-frequency images show power suppression slowly increasing for AR2\_3 and decreasing for AR2\_4, suggesting that the rate of change of the intensity is related to the rate of change of the oscillatory power. Again the lower frequency bands below 3 mHz show higher power levels than the bands above 3 mHz.  The relatively low signal-to-noise ratio in these panels can be increased by enlarging the size of the selected subregions.

\par
Region AR2\_5 (Figure~\ref{figure10}) is a quiet region that provides an observation of the acoustic power characteristics outside flaring regions. This control region shows  a constant oscillatory power with no systematic temporal changes within the frequency bands (Figure~\ref{figure11} (c)). While it does show higher power at lower frequencies, the enhancement is much lower than that seen in the flare regions.  Quasi-periodic fluctuations are also seen with a periods of $\sim$7 minutes in the  1--2 mHz band and $\sim$1 minute above 5.0 mHz.  However, these may be a result of a low signal-to-noise ratio.

%----------------------------------------------------------------------------------------
%	FIGURE 11
%----------------------------------------------------------------------------------------
 \FloatBarrier     
%----------------------------------------------------------------------------------------

%----------------------------------------------------------------------------------------
%	DISCUSSION & CONCLUSIONS
%----------------------------------------------------------------------------------------     
\newpage      

\section{DISCUSSION AND CONCLUSIONS\label{sec:discussion}}         
\par
The results in this paper demonstrate that H$\alpha$ observations of chromospheric oscillations in the p-mode band can provide information about the physical processes occurring in flaring regions. In particular, variations in the oscillatory power as a function of frequency, spatial position, and time can be used to probe energy transport at different heights within a flare.

\par
Figures~\ref{figure5} and \ref{figure9} show a suppression of power that first migrates in time from high to low frequencies in a flare, with a subsequent restoration of the power starting at low frequencies and progressing back to high frequencies. This produces a ``V''-shaped feature in the images of the power as a function of frequency and time.  The shape can be understood as a consequence of the nature of the observations, the behavior of the H$\alpha$ spectral line during a flare, and the height dependence of the frequency of the maximum oscillation amplitude.  Wang \emph{et al.} \citeyearpar{Wang2000}, with their high-cadence H$\alpha$ observations from the Big Bear Solar Observatory, found high-frequency fluctuations that correlate with HXR elementary bursts.  These hard X-ray emissions could be signatures of sites of fine structures where individual magnetic reconnection processes are taking place \citep{Wang2000}.

\par
One of the earliest results of studies of solar oscillations showed that the frequency at which the waves reach their maximum amplitude increases with height in the solar atmosphere \citep{Noyes1962}. In the photosphere, the maximum amplitude occurs at periods around five minutes (frequency near 3.3 mHz), while in the chromosphere the maximum occurs at periods of three minutes (frequency near 5.5 mHz). It is also known that the wings of a spectral line are formed at lower heights in the solar atmosphere than the core of the line, thus observations of oscillations obtained in the wing of a spectral line will be dominated by lower-frequency power than observations in the core of the line.  

\par
The observations discussed here are obtained with a filter that is centered on the wavelength of the H$\alpha$ line core in the quiet Sun. During the course of the flare, the motion of the plasma will change the wavelength of the line due to the Doppler effect, so that the filter bandpass will admit a higher proportion of light from the wings of the spectral line rather than from the core. Since the wings of the line are formed at lower heights in the solar atmosphere, and since the peak amplitude of the p-modes occurs at lower frequencies at lower heights, the net effect is to reduce power at high frequencies. This reduction moves to lower frequencies as the flare progresses and the spectral line is increasingly Doppler-shifted. The overall observed intensity also increases as the brighter wings contribute a larger portion of the signal. As the flare energy decreases, the solar plasma motions die out and the spectral line core moves back toward the center of the filter bandpass, restoring the visibility of the high-frequency power and decreasing the overall intensity. A plasma velocity of 5 km s\textsuperscript{-1}, easily created in a flare, would move the line core by 0.1\AA\ , which is a substantial fraction of the 0.6\AA\ bandpass of the GONG filter.

\par
The dark lanes in wave power, also observed by Jackiewiez and Balasubramaniam \citeyearpar{Jackiewicz2013}, could be where the magnetic field absorbs or scatters the acoustic waves.  In the photosphere, sunspots are known to be areas of suppressed acoustic mode power \citep{Braun1987}.  The appearance of the dark lanes depends on frequency, as seen in Figure~\ref{figure3}, which may provide information on the structure of the magnetic field as well as aspects of energy transport during the flare.  In addition, the time-frequency maps of subregions located on dark lanes show diverse behavior (e.g. AR3\_3 and AR3\_2) further suggesting that there is a variation in the underlying magnetic field.  Several deductions can be made about the dark appearance of the lanes, which could indicate that energy is being removed from the observed wave frequencies and perhaps converted into the thermal energy of the flare, or scattered into other wave modes with frequencies higher than 8 mHz, or absorbed by the magnetic field in the flare, or dampened by magnetic reconnection.  We believe that the correct explanation is that the wave energy is being converted into thermal energy, due to the simultaneous increase in both \emph{GOES} X-ray flux and H$\alpha$ intensity.  These possibilities can be investigated by applying the PMM technique to simultaneous magnetograms acquired by GONG.  Furthermore, there is a trade-off between signal-to-noise ratio and spatial resolution. Larger subregions increase the signal-to-noise ratio but decrease the spatial resolution. In this paper we chose to have higher spatial resolution in order to investigate oscillations within the narrow dark lanes.

\par
Generally, there is a tendency for an excess of power at low frequencies below 2 mHz compared to higher frequencies. This excess can be as much a factor of 30 for the regions in Figure~\ref{figure5}, but it is also present in the quiet region in Figure~\ref{figure11} at a much lower level (a factor of about 4). Since these are ground-based intensity observations, it is quite possible that some of this excess is caused by fluctuations in the Earth's atmospheric transparency. However, the marked increase in the flaring regions suggests that low-frequency power is enhanced during a flare. If this low-frequency power excess is a feature of strong flares, it may arise from an instability in the chromosphere and provide an early warning of the flare onset.

\par
This pilot project demonstrates that the application of PMMs to H$\alpha$ intensity observations opens up a number of new avenues to explore the physical processes in flares. The temporal and spatial variations of acoustic wave power show intriguing features that contain information about the energy transport and magnetic field variations as a function of height within flaring regions. 

\par
There are several paths to follow that will further develop the method. The most informative step is the comparison of the results with the GONG magnetograms. The correlation of changes in oscillatory power with the characteristics of the magnetic field should provide additional information on the underlying physical processes. In addition, the PMM method can also be applied to the magnetograms since they are simultaneously observed at a cadence of once per minute. Additional steps will be the analysis of larger subregions, extension of the PMM method to earlier times to search for flare precursors; application of the method to longer time spans to increase the frequency resolution, analysis of additional cases of strong flares, and accumulation of statistically significant measurements.  Eventually, we may be able to construct a new picture of the physics of a flare based on its acoustic signatures.

%----------------------------------------------------------------------------------------
%	ACKNOWLEDGEMENTS
%----------------------------------------------------------------------------------------  

\acknowledgements
The authors acknowledge the advice given by Dr.\ Jason Jackiewicz and Dr.\ Nathan De Lee.  This work utilized data from the GONG H$\alpha$ network, operated by The National Solar Observatory (NSO) and The Association of Universities for Research in Astronomy (AURA), and which was originally commissioned by The Air Force Weather Agency (AFWA).  T.M. acknowledges support from the Fisk-Vanderbilt Masters-to-PhD Bridge Program, including specifically funding support through NSF PAARE grant AST-1358862 and a Harriett Jenkins Graduate Fellowship from NASA.

%----------------------------------------------------------------------------------------
%	REFERENCE LIST
%----------------------------------------------------------------------------------------

%\input{\AJPAPER1.bbl}
%\bibliography{solarbib}

\end{document}